\documentclass[12pt]{iopart}

\usepackage{graphicx}
\usepackage{amssymb}
\usepackage{subfig,color}

\usepackage{iopams}  
\begin{document}

\title{Inertial and topological effects on a 2D electron gas}

\author{J. Brand\~ao $^{1,2}$, C. Filgueiras $^3$, Moises Rojas $^3$, F. Moraes $^{1,4}$}

\address{$^1$Departamento de F\'{\i}sica, CCEN,  Universidade Federal da Para\'{\i}ba, Caixa Postal 5008,  58051--900, Jo\~ao Pessoa, PB, Brazil}
\address{$^2$Instituto Federal de Educa\c{c}\~ao, Ci\^encia e Tecnologia do Sert\~ao Pernambucano, Campus Salgueiro, 56000-000, Salgueiro, PE, Brazil}
\address{$^3$Departamento de F\'{\i}sica (DFI),
Universidade Federal de Lavras (UFLA), Caixa Postal 3037,
37200--000, Lavras, Minas Gerais, Brazil}
\address{$^4$Departamento de F\'{\i}sica, Universidade Federal Rural de Pernambuco, 52171--900, Recife, PE, Brazil}

\ead{juliobrandaosilva@gmail.com (Julio Brand\~ao),
cleversonfilgueiras@yahoo.com.br (Cleverson Filgueiras),
moises.leyva@dfi.ufla.br (Moises Rojas), fernando.jsmoraes@ufrpe.br (Fernando Moraes)}

\date{\today}

\begin{abstract}
In this work, we study how the combination of rotation and a topological defect can  influence the energy spectrum of a two dimensional electron gas in a strong perpendicular magnetic field. A deviation from the linear behavior of the energy as a function of magnetic field, caused by a tripartite term of the Hamiltonian, involving magnetic field, the topological charge of the defect and the rotation frequency,  leads to novel features which include a range of magnetic field without corresponding Landau levels and changes in the  Hall quantization steps.
\end{abstract}

\pacs{{73.40.-c, 61.72.Lk, 73.43.Cd}}

\vspace{2pc}
\noindent{\it Keywords}: Integer Quantum Hall Effect, Topological Defects, Inertial Effects\\
%
\submitto{\JPOC}
%
%
%

\section{Introduction}

When a two-dimensional electron gas (2DEG) is submitted to a perpendicular magnetic field, the free electron energy levels become quantized into degenerate Landau levels (LL) \cite{RevModPhys.54.437}. This degeneracy, as well as the splitting of the levels,  increases linearly with the magnetic field. Disorder broadens the levels into bands with extended states in the central region and localized states  in the tail regions. This leads to a plethora of phenomena, among them, the Integer Quantum Hall effect (IQHE) \cite{prange1987quantum}, which may be affected by geometric and inertial factors. For instance, curvature is known to affect the IQHE \cite{Bulaev2003180,Poux2014,PhysRevLett.117.266803}. In fact, curvature may be used as a tool to manipulate the electronic structure of charge carriers confined to a surface \cite{santos2016geometric,Liu2017123,Liang2016246}. Rotation, as well, has its effects on the IQHE as we reported in a previous publication \cite{0295-5075-110-2-27003}. The aim of this article is to further elucidate the influence of curvature and rotation on the LL and on the IQHE. This is done with the inclusion of a topological defect (disclination) and  by making the system rotate around the defect axis. Separately, the disclination \cite{deLima2012} and the rotation \cite{0295-5075-110-2-27003,Brando} introduce  subtle modifications in the LL spectrum without  changing the linear dependence of the energy on the magnetic field. Together, they make the energy dependence on the magnetic field become parabolic and there appears a range of magnetic field free of LL, as  presented below.

There is some similarity between the application of a magnetic field and rotation on free charged particles \cite{johnson:649,0295-5075-54-4-502}. It comes mainly from the Coriolis force which, if it could act alone, would work like the magnetic force, creating  analogue LL. However, its companion, the centrifugal force, breaks the degeneracy and the sample-length independence of the LL. In fact, the Coriolis and centrifugal contributions to the quantum Hamiltonian together lead  to a coupling between the particle total angular  momentum and the rotation.  Therefore, the  combination of magnetic field and rotation provokes a shift and a splitting of the original LL  due to this coupling \cite{Brando}.

We consider a non-interacting free electron gas in a rotating disk with a disclination at its rotation axis, under the influence of a perpendicular, uniform, magnetic field.  Charged particles in a rotating Hall sample  with a magnetic field were already studied in Ref. \cite{Brando}, where it was pointed out that  rotation breaks the degeneracy of the LL.  Furthermore, the counting of states fully occupied below the Fermi energy may change, altering the Hall quantization steps. Charged particles in a  sample with a single disclination, in the presence of an orthogonal constant magnetic field, were also studied in Ref. \cite{deLima2012}, where it was found a modification in the Hall conductivity steps due to the defect. Our purpose here is to investigate  the combined influence of  rotation and disclination  in the Hall conductivity. As expected, and in fact verified in Refs. \cite{Brando,deLima2012}, we find that   rotation and the disclination, separately couple to angular momentum, as does the magnetic field. But, when rotation, disclination, and magnetic field act together on the free electron gas, a new coupling is found involving all of them simultaneously. This makes the LL as function of the magnetic field to bend from the usual straight lines and introduces a region of magnetic field without LL, as shown below.
\section{Energy levels}
In three dimensions, a disclination is a topological defect \cite{1063-7869-48-7-R02} associated to the removal of a wedge of material with the subsequent identification of the loose ends (Volterra process), introducing an angular deficit which changes the boundary condition on the angular variable from $\phi \rightarrow \phi + 2\pi$ into $\phi \rightarrow \phi + 2\pi\alpha$. Here, $\alpha <1$ expresses the removed wedge  angle of $2\pi(1-\alpha)$. Conversely, if a wedge is added, $\alpha >1$. Effectively, the new boundary condition can be applied by working in a  background space with the line element
\begin{equation}
ds^2 = dr^2 + \alpha^2 r^2 d\phi^2 +dz^2. \label{metric}
\end{equation}
As shown in  \cite{1063-7869-48-7-R02}, the Frank vector which characterizes a disclination is nothing more than the curvature flux associated to the defect. Since the above line element corresponds to a  curvature scalar given by $R= 2\left( \frac{1-\alpha}{\alpha}\right) \frac{\delta(\rho)}{\rho}$, as shown in \cite{doi:10.1142/S0217732309029995}, its flux is therefore 
\begin{equation}
\oint R \rho d\rho d\phi= 4\pi \left( \frac{1-\alpha}{\alpha}\right) = F, \label{frank}
\end{equation}
giving the Frank vector $F$, or topological charge of the disclination. This result still holds for a two-dimensional surface with a disclination, which is the subject of this article.

The disclination introduces anisotropy which is manifest in the effective mass approximation. In this approach, the bottom of the conduction band of a isotropic material is approximated by a parabola corresponding to the dispersion relation $E(\vec{k})= \frac{\hbar^2}{2m^*}k^2$, where $m^*$ is the effective mass. For anisotropic materials, this becomes $E(\vec{k})= \sum_{i,j} \frac{\hbar^2}{2m_{ij}^*}k_i k_j$. The introduction of a disclination into an otherwise isotropic material breaks the global rotational symmetry of the system, reducing it to rotations around the disclination axis. The immediate consequence of this is the transformation of the inverse isotropic effective mass from a scalar to a tensor:
\begin{equation}
\frac{1}{m^*}= \left(
\begin{array}{ccc}
 1/m_r & 0 & 0 \\
 0 & 1/m_{\phi} & 0\\
 0 & 0 & 1/m_z 
\end{array} \right),
\end{equation}
where $(r,\phi,z)$ are the usual cylindrical coordinates. The kinetic energy operator then becomes 
\begin{equation}
K= -\frac{\hbar^2}{2}\left[\frac{1}{m_r}\left(\frac{\partial^2}{\partial r^2} + \frac{1}{r}\frac{\partial}{\partial r}\right)+ \frac{1}{m_{\phi}r^2}\frac{\partial^2}{\partial \phi^2} +  \frac{1}{m_{z}}\frac{\partial^2}{\partial z^2}\right]
\end{equation} which, for a two-dimensional system can be written as 
\begin{equation} 
K= -\frac{\hbar^2}{2m_r}\left[\frac{\partial^2}{\partial r^2} + \frac{1}{r}\frac{\partial}{\partial r} + \frac{1}{(\alpha ')^2r^2}\frac{\partial^2}{\partial \phi^2}\right], \label{K}
\end{equation}
where $\alpha '=m_{\phi}/m_r$. Of course $\alpha$ and $\alpha '$ are not necessarily the same but the existence of the former implies in the appearance of the latter in the kinetic energy operator. The term between square brackets is the Laplacian in the effective geometry associated to the line element given by Eq. (\ref{metric}) with $\alpha$ replaced by $\alpha '$. This being clear, on what follows, we drop the prime from $\alpha '$ and the subscript $r$ from $m_r$ for simplicity of notation.

As  discussed in  \cite{Brando}, the Hamiltonian in cylindrical coordinates of a particle in a rotating disk, in the presence of a magnetic field   $\vec{B}=B \hat{z}$, can be written as
\begin{eqnarray}
H=\frac{[\vec{p}-q\vec{A}-m(\vec{\Omega}\times\vec{r})]^2}{2m}-\frac{m(\vec{\Omega}\times\vec{r})^2}{2}+qV\; ,\label{hamiltonianageral}
\end{eqnarray}
where $V$ and $\vec{A}$ are the scalar and vector electromagnetic potentials,  given by
\begin{eqnarray}
V=-\frac{\Omega Br^2}{2},\\
\vec{A}=(0,\frac{Br}{2\alpha},0),
\end{eqnarray}
and  $\vec{\Omega}=\Omega\hat{z}$ is the angular velocity of the disk. The electric field associated to the scalar potential appears from the transformation of the applied magnetic field to the rotating frame. The disclination factor, $\alpha$, appearing in the vector potential compensates the change $\phi \rightarrow  \alpha \phi$ giving the correct value of the magnetic flux through a circle of radius $r$ in the plane. That is, $\oint A_{\phi} ds= \oint \frac{Br}{2\alpha}r \alpha d\phi = \pi r^2 B$, where $ds = r \alpha d\phi$ comes from the metric (\ref{metric}) with $z=const.$ and $r=const.$, which define the circle.
Again, due to the change  $\phi \rightarrow  \alpha \phi$, we write the angular momentum operator as $\vec{p}_\phi=-\hat{\phi}\frac{i\hbar}{\alpha r}\frac{\partial}{\partial\phi}$.
Thus, the Hamiltonian can be written as
\begin{eqnarray}
H=\frac{p^2}{2m}-\mu rp_\phi+\beta r^2 \; , \label{H6}
\end{eqnarray}
with
\begin{eqnarray}
\mu&=&\frac{qB}{2m\alpha}+\Omega \label{mu}
\end{eqnarray}
and
\begin{eqnarray}
\beta&=&\frac{q^2B^2}{8m\alpha^2}+\frac{qB\Omega F}{8\pi}.\label{beta}
\end{eqnarray}
The $\mu rp_\phi$ term in Eq. (\ref{H6}) contains the usual coupling between the magnetic field and the angular momentum  plus the coupling between rotation and angular momentum (see Eq. (\ref{mu})). The quadratic term in Eq. (\ref{H6}) has the usual  contribution from the magnetic field and a term which couples magnetic field $B$, rotation speed $\Omega$ and the topological charge $F$ of the defect   (see Eq. (\ref{beta})).

For this Hamiltonian, the Schr\"odinger equation can be written as
\begin{eqnarray}
-\frac{\hbar^2}{2m}\nabla^2\psi+i\frac{\mu\hbar}{\alpha}\frac{\partial\psi}{\partial\phi}+\beta r^2\psi=E\psi, \label{eqschrodinger}
\end{eqnarray}
where the Laplacian operator (in fact the Laplace-Beltrami operator) is given by $\nabla^2= \frac{\partial^2}{\partial r^2} + \frac{1}{r}\frac{\partial}{\partial r}+ \frac{1}{\alpha^2 r^2}\frac{\partial^2}{\partial \phi^2} $ (see Eq. (\ref{K}).
With the \textit{ansatz} $\psi=R(r)e^{-il\phi}$ , Eq. (\ref{eqschrodinger}) becomes
\begin{eqnarray}
r^2R''+rR'+(-\sigma^2r^4+\gamma^2 r^2-\frac{\ell^2}{\alpha^2})R=0, \label{eqemr}
\end{eqnarray}
where 
\begin{eqnarray}
\sigma^2=\frac{q^2B^2}{4\hbar^2\alpha^2}+\frac{mqB\Omega}{\hbar^2}\left(\frac{1-\alpha}{\alpha}\right)
\end{eqnarray}
and
\begin{eqnarray}
\gamma^2=\frac{2m}{\hbar}\left(\frac{E}{\hbar}-\frac{qB\ell}{2m\alpha^2}-\frac{\Omega \ell}{\alpha}\right).
\end{eqnarray} 
Writing $\sigma r^2=\xi $ and looking at the asymptotic limit as $\xi\rightarrow\infty$, the general solution to this equation will be given in terms of  $\rm M(a,b,\xi)$, 
 the {\it confluent hypergeometric function of the first kind}  \cite{nist},
{\small\begin{eqnarray}
R\equiv R_{\ell}\left( \xi \right)=a_{\ell}e^{-\frac{\xi}{2}}\xi^{\frac{|\ell|}{2\alpha}}{\rm M}\left(\frac{-\gamma^{2}}{4\sigma}+
\frac{|\ell|}{2\alpha}+\frac{1}{2} ,1+\frac{|\ell|}{\alpha}
,\xi\right)
\nonumber\\
+b_{\ell}e^{-\frac{\xi}{2}}\xi^{-\frac{|\ell|}{2\alpha}}{\rm M}
\left(\frac{-\gamma^{2}}{4\sigma}-\frac{|\ell|}{2\alpha}+\frac{1}{2} ,1-\frac{|\ell|}{\alpha}
,\xi\right).
\label{general_sol_2_HO}
\end{eqnarray}}
In Eq. (\ref{general_sol_2_HO}), $a_{\ell}$ and $b_{\ell}$ are, respectively, the coefficients of the {\it regular} and {\it irregular}
solutions. Notice that the term {\it irregular} stems from the fact that it diverges as $\xi\rightarrow0$. 
We consider this case since the disclination, described by a cone-like background, may introduce a singular
curvature potential in the problem \cite{Jensen2011448}.
As showed in \cite{lasersingular} in the study of the emission of X-ray laser beams from metals, singularities have important impact in quantum systems. In dealing with singularities, extra care is needed since,  when they occur, the Hamiltonian may not be not self-adjoint. For the specific case of the conical singularity associated to a disclination, Ref. \cite{filgueiras2008quantum} gives the prescription to make a Hamiltonian self-adjoint by extending its domain. This leads to the inclusion of the \textit{irregular} term in Eq. (\ref{general_sol_2_HO}). The \textit{regular} solution  corresponds to the case where the core of the disclination which carries the curvature singularity is artificially removed (not considered). For the sake of comparison we present our results bellow for both cases: \textit{regular} (core removed) and \textit{total} (core included, meaning \textit{regular} + \textit{irregular}).

In order to have a finite polynomial function (the hypergeometric series has to be convergent in order to have a physical solution), the condition ${{\rm a}=-n}$, where $n$ is a positive integer number,  has to be satisfied. From this condition, the  possible values for the energy are given by
\begin{eqnarray}
E&=&\frac{\hbar \omega_c \ell}{2 \alpha^2}+\frac{\hbar\Omega\ell}{\alpha}+\nonumber\\  & &\hbar\sqrt{\frac{\omega_{c}^2}{\alpha^2}+4\omega_c\Omega\frac{\left(1-\alpha\right)}{\alpha}}\left[n\pm\frac{|\ell |}{2\alpha}+\frac{1}{2}\right],\label{Energyspectrum}
\end{eqnarray}
where $\omega_c=qB/m$ is the cyclotron frequency. It should be noted that, for the case where the singular interaction is absent, 
only the regular solutions contribute for the bound state wave function ($b_{\ell}\equiv0$), and the energy is given by Eq. (\ref{Energyspectrum}) with the plus sign together with the condition $| \frac{\ell}{\alpha}| \geq 1$. On the other hand, if a singular interaction is present, then the energy levels with the minus sign  exist but the constraint $| \frac{\ell}{\alpha} | <1$ must be observed since this is the condition for the irregular wave function to be square integrable \cite{Azevedo2015196}. 
The energy spectrum above must be analyzed in terms of the values  the parameter $\alpha$ can assume, since the condition 
$| \frac{\ell}{\alpha}| \geq 1$($|\frac{\ell}{\alpha} | <1$) for the
regular (irregular) solution has to be fulfilled.  In summary, in the absence of the coupling between the wavefunctions and the singular curvature, the constraint  $| \frac{\ell}{\alpha}| \geq 1$ must be imposed, which restricts the allowed momentum eigenvalues $\ell$. This guarantees that such wave functions are regular as $r\rightarrow 0$. If such coupling is present, then the particles are allowed to have other values of the  angular momentum eigenvalues $\ell$ without the constraint. We will examine these two situations in order to show how singular effects are important when they appear in the system.

We remark that the fields $B$ and $\Omega$ are externally tunable parameters. Now, considering $\alpha=1$ and $\Omega\neq 0$, \textit{i.e.}, when the system rotates but has no disclination, we recover the result found in Ref. \cite{Brando}. On the other hand, by considering $\alpha\neq 1$ and $\Omega=0$, we recover the result presented in the Refs. \cite{deLima2012}, \cite{furtado}. The disclination alone introduces a change in the angular momentum that breaks the infinite degeneracy of the Landau levels \cite{furtado}. The presence of rotation by itself breaks the degeneracy between states with opposite angular momentum and introduces an energy shift depending on the rotation  \cite{Brando}. But, when we have magnetic field, disclination and rotation simultaneously, there exists a term in the energy, $4\omega_c\Omega\frac{\left(1-\alpha\right)}{\alpha}$, that represents the coupling of these three elements, \textit{i.e.}, if either  $\Omega=0$, $\alpha=1$ or $B=0$ this term vanishes from the energy expression. 

In the absence of magnetic field,  the nature of the wave function is completely changed as seen below. Considering  $B=0$, Eq. (\ref{eqemr}) becomes
%
\begin{eqnarray}
r^2R''+rR'+(\gamma^2 r^2-\frac{\ell^2}{\alpha^2})R=0, \label{eqemrwithoutb}
\end{eqnarray}
which is a Bessel equation with $\gamma^2=\frac{2m}{\hbar}\left(\frac{E}{\hbar}-\frac{\Omega l}{\alpha}\right)$. Then the energy spectrum can be written as 
\begin{eqnarray}
E=\frac{\hbar^2\gamma^2}{2m	}+\frac{\hbar\Omega \ell}{\alpha}, \label{Ejohn}
\end{eqnarray}
and the wave function is
\begin{eqnarray}
\psi=J_{\ell/\alpha}(|\gamma|r) e^{-\frac{i\ell\theta}{\alpha}}, \label{Pjohn}
\end{eqnarray}
where $\gamma$ is a continuous variable if the sample is infinite. For a finite size sample, $\gamma$ is discrete and is found by the condition $J_{\ell/\alpha}(|\gamma|a)=0$, where $a$ is the radius of the sample.  Therefore, the presence of the disclination modifies the result  obtained by Johnson in Ref.  \cite{johnson:649}, who studied the effect of rotation on the energy spectrum of  free charges and its consequences to the Hall effect.  Effectively, we obtain a rescaling of the angular momentum $\ell$ by the defect parameter $\alpha$ as seen in Eqs. (\ref{Ejohn}-\ref{Pjohn}).   For $\alpha=1$ we recover the result of Johnson\cite{johnson:649}.

\section{Electronic structure}
\begin{figure*}[!htp]
\centering
    \subfloat{\includegraphics[width=0.3\linewidth]{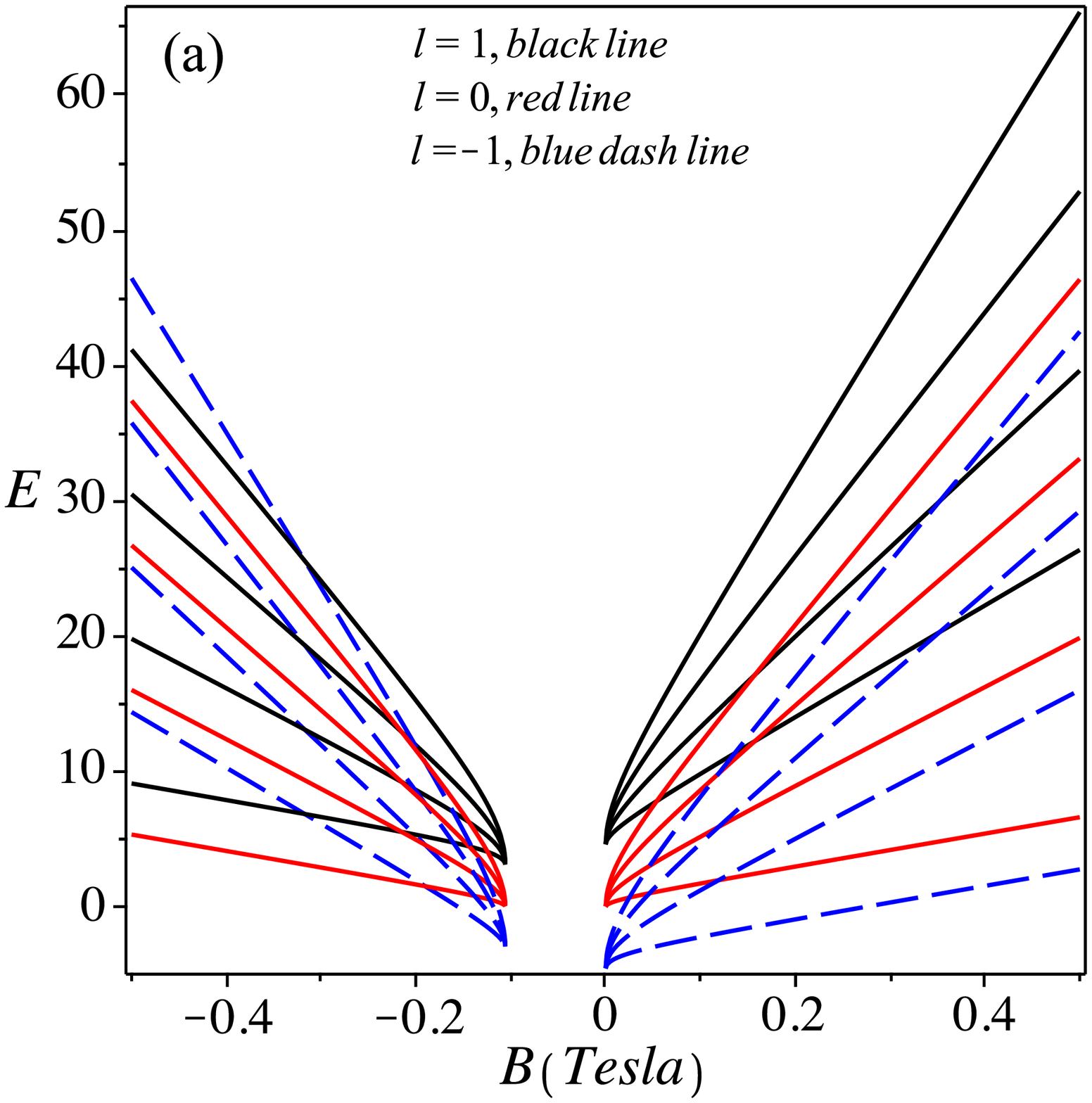}}
    \subfloat{\includegraphics[width=0.3\linewidth]{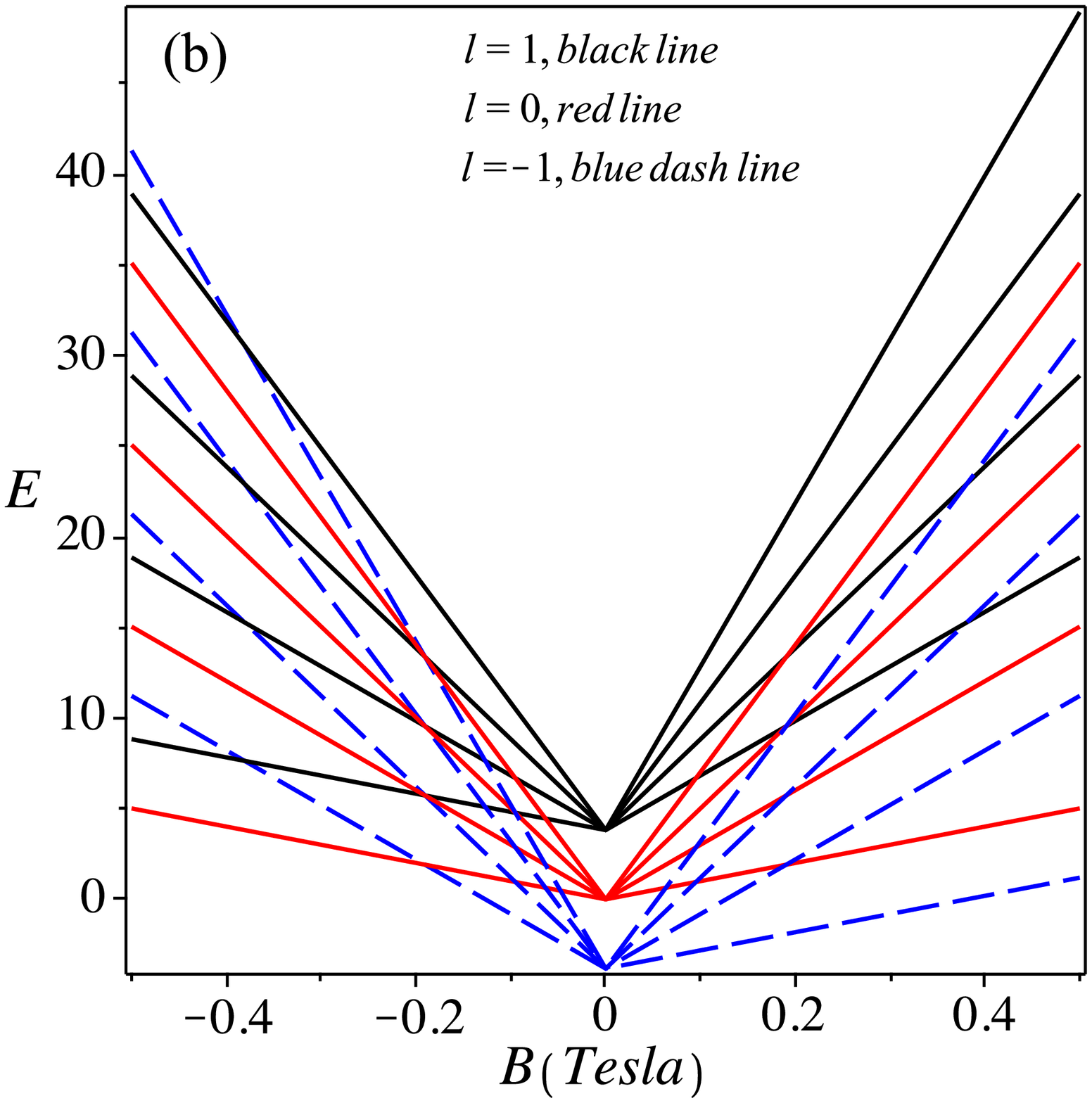}}
    \subfloat{\includegraphics[width=0.3\linewidth]{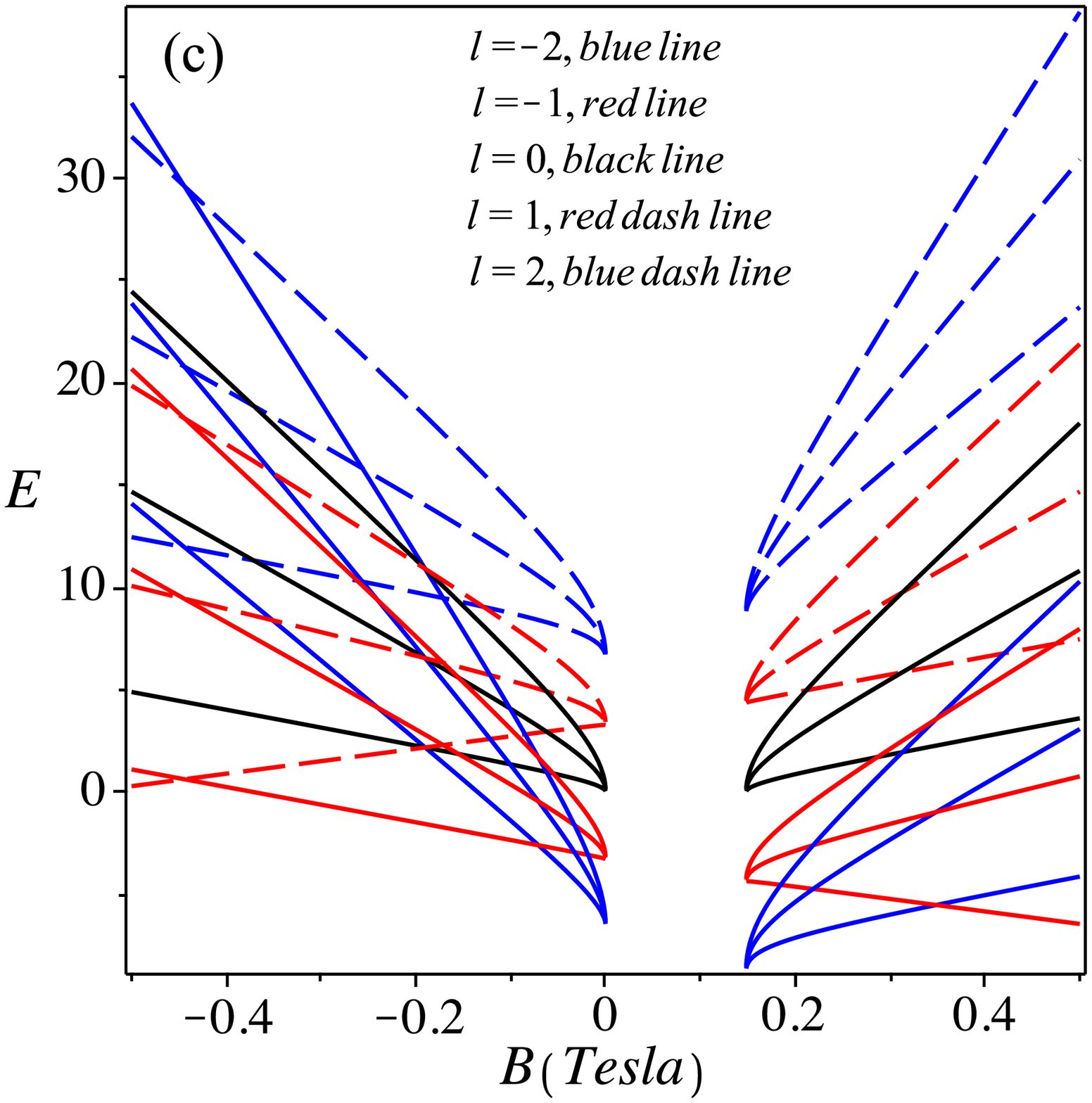}}
    \caption{\small Energy levels, in temperature units (K), versus magnetic field for the first few values of $n$ . We consider a rotating disk (500 GHz) with a disclination for {\bf (a)} $\alpha=5/6$  (positive curvature), {\bf (b)} $\alpha=1$(flat) and {\bf (c)} $\alpha=7/6$ (negative curvature). Notice the gap appearing in the  magnetic field range.} 
          \label{figniveis}
\end{figure*}

The energy levels of a rotating disk with a disclination in the presence of a magnetic field are given  by Eq. (\ref{Energyspectrum}), whose representation can be seen in  Fig. \ref{figniveis}. We used $\alpha = 5/6$ and $\alpha=7/6$, realistic values for the hexagonal lattice, which correspond to  positive and  negative curvature defects, respectively.  The values of the quantum number $\ell$, considering the irregular part of the spectrum, are indicated in the figures. As expected \cite{Brando}, the presence of rotation splits the levels. When there is also a disclination in the system, there appears a range of magnetic field without corresponding bound energy states (LL). This happens when the square root in Eq. (\ref{Energyspectrum}) becomes imaginary for $\frac{\omega_{c}^2}{\alpha^2}+4\omega_c\Omega\frac{\left(1-\alpha\right)}{\alpha} <0$, which may be obtained by either inverting the direction of rotation or the magnetic field. Or else, by replacing a sample with a positive curvature  disclination ($\alpha <1$) by another one with a negative curvature disclination ($\alpha >1$), which is what we present in Fig. \ref{figniveis}.

The term $4\omega_c\Omega\frac{\left(1-\alpha\right)}{\alpha}$  exists only when  magnetic field, rotation and a disclination occur simultaneously in the sample.  This explains why there is no bound state (LL) in the  magnetic field range $-\frac{4m\Omega}{q}\alpha (1-\alpha )<B<0$ if $\alpha <1$  or  $0<B<\frac{4m\Omega}{q}\alpha (\alpha -1 )$ if $\alpha >1$.  Notice that this range depends on the material through the effective electron mass, $m$.  In summary, there is a competition between  the three elements (B, $\Omega$ and $\alpha$)  which affects directly the energy spectrum. The  scattering states for $B$ in the  aforementioned range is due to the fact that, in this range, the magnetic field is not strong enough to overcome the combined effect of rotation and disclination. Since the gap in the magnetic field range, shown in Fig. \ref{figniveis}, destroys the Landau levels, rotation (plus disclination) can be used to bring the Hall effect to the classical regime, where many applications as Hall sensors abound. 

Without the term term $4\omega_c\Omega\frac{\left(1-\alpha\right)}{\alpha}$, the energy is linear in $B$ as easily seen in Eq. (\ref{Energyspectrum}). The presence of this term distorts the straight lines into the parabolas shown in Fig. \ref{figniveis}. Also, in Fig. \ref{figniveis}, it is  clear the splitting of levels due to rotation, and how the disclination mediates this effect. 

\section{Hall conductivity}

At zero temperature and considering that the Fermi energy $E_F$ is in an energy gap, the Hall conductivity can be written as \cite{Streda}
\begin{equation}
\sigma_H(E_F,0)=\frac{e}{S}\frac{\partial N}{\partial B} \; ,
\end{equation}
where $N$ is the number of states below the Fermi energy and $S$ is the area of the surface. The density of states is given by
\begin{equation}
n(E)=\frac{|eB|}{2\pi \hbar}\sum_{n,l}\delta(E-E_{n,l}) \; . 
\end{equation}
Therefore, one can obtain $N$, as
\begin{equation}
N=S\int_{-\infty}^{E_F}n(E)dE=\frac{S|eB|}{h}  n_0,
\end{equation}
where $n_0=(n+1) n_\ell$ is the number of fully occupied LLs below $E_F$. The number of occupied $\ell$ states is represented by $n_\ell$. The Hall conductivity is then
\begin{equation}
\sigma_H(E_F,0)=-\frac{e^2}{h}n_o \; .
\end{equation}

For $T\neq0$, we consider the expression for the Hall conductivity obtained in the clean limit (absence of impurities) given by\cite{graphenecondutivity} 
{\small\begin{eqnarray}
 \sigma_{H}(\mu,T)=\int_{-\infty}^{\infty}\left(-\frac{\partial f_0}{\partial E}\right)\sigma_{H}(E,0)dE\label{conduc}
\end{eqnarray}}
where $\mu$ is the chemical potential, $T$ is the temperature and $f_0$ is the Fermi-Dirac distribution. We  express the energy scale in  units of temperature. We consider the parameters of a GaAS: $m=0,067m_e$, where $m_e=9,11\times10^{-28}g$ is the electron rest mass.

\begin{figure*}[!htp]
\centering
    {\bf a}\subfloat{\includegraphics[width=0.40\linewidth]{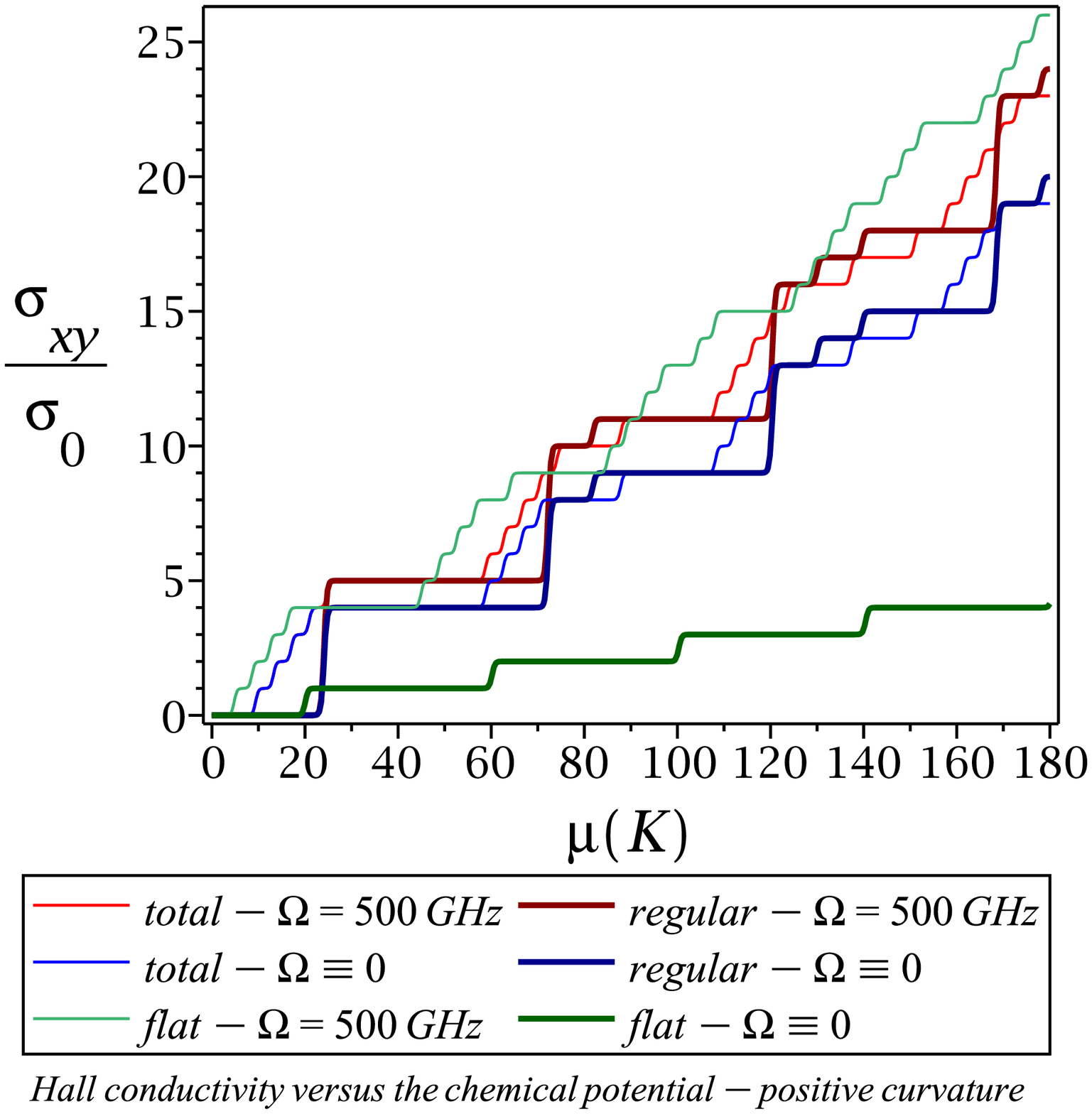}}
    {\bf b}\subfloat{\includegraphics[width=0.40\linewidth]{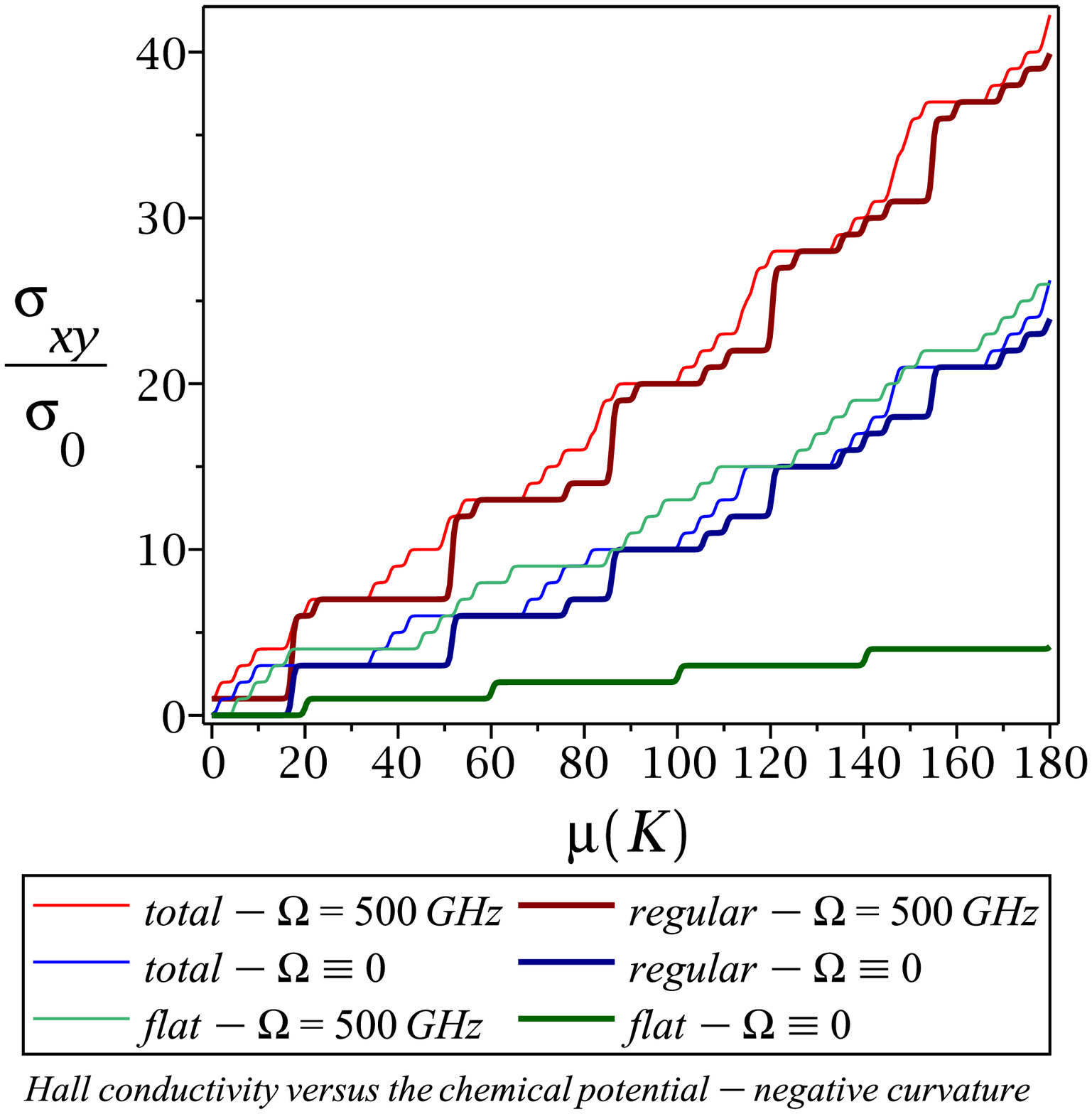}}
\caption{\small Hall conductivity versus the chemical potential, in temperature units (K), for a rotating disk (500GHz) with disclination: {\bf (a) } $\alpha = 5/6$ and {\bf (b)} $\alpha = 7/6$. For comparison we include  the Hall steps for the stationary ($\Omega=0$) and flat  (no disclination), cases. We consider $T=0.3{\rm K}$, $B=2T$ and $\sigma_0\equiv -e^{2}/h$.} 
          \label{chemical}
\end{figure*}

\begin{figure*}[!htp]
\centering
    {\bf a}\subfloat{\includegraphics[width=0.40\linewidth]{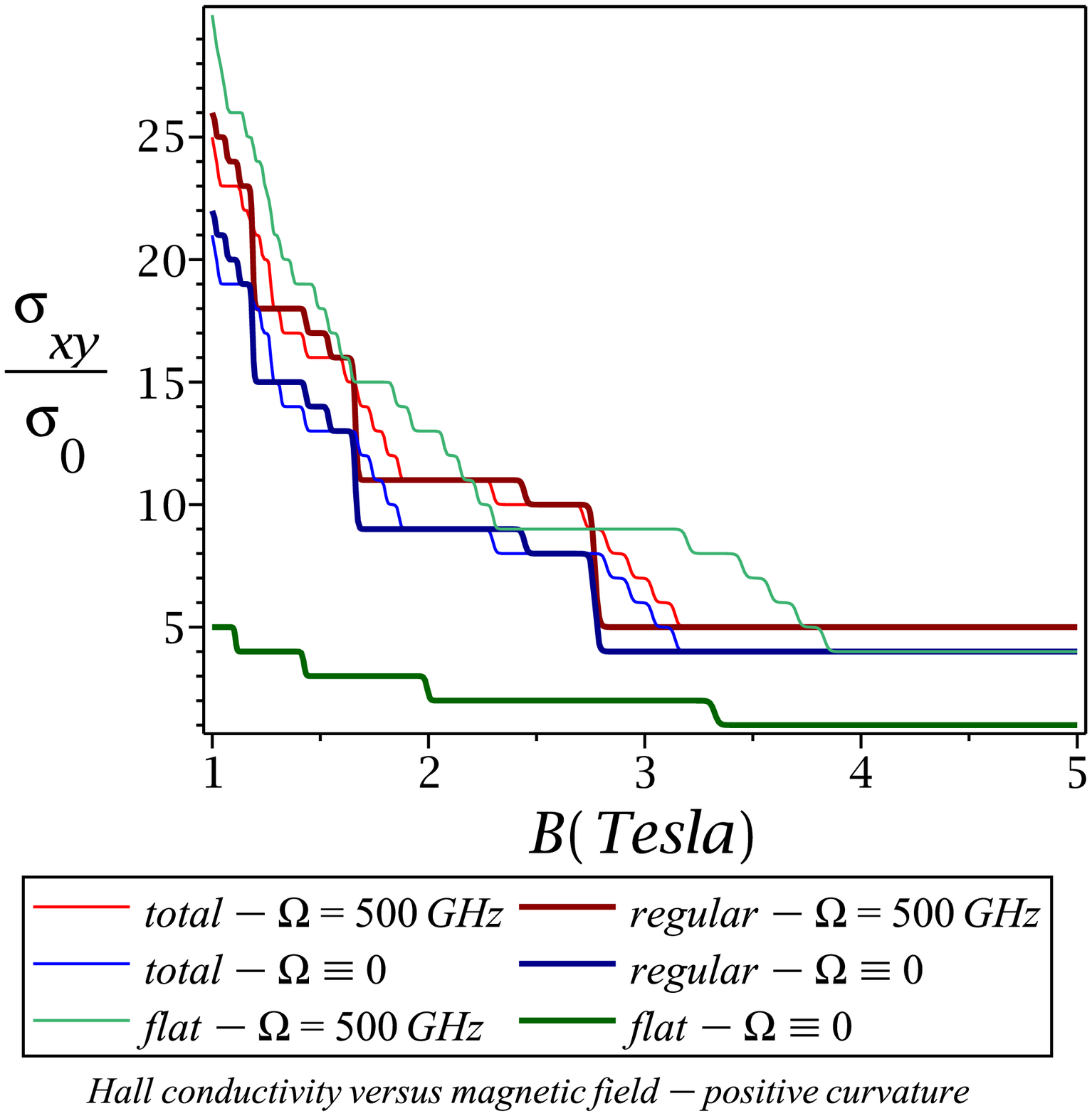}}
    {\bf b}\subfloat{\includegraphics[width=0.40\linewidth]{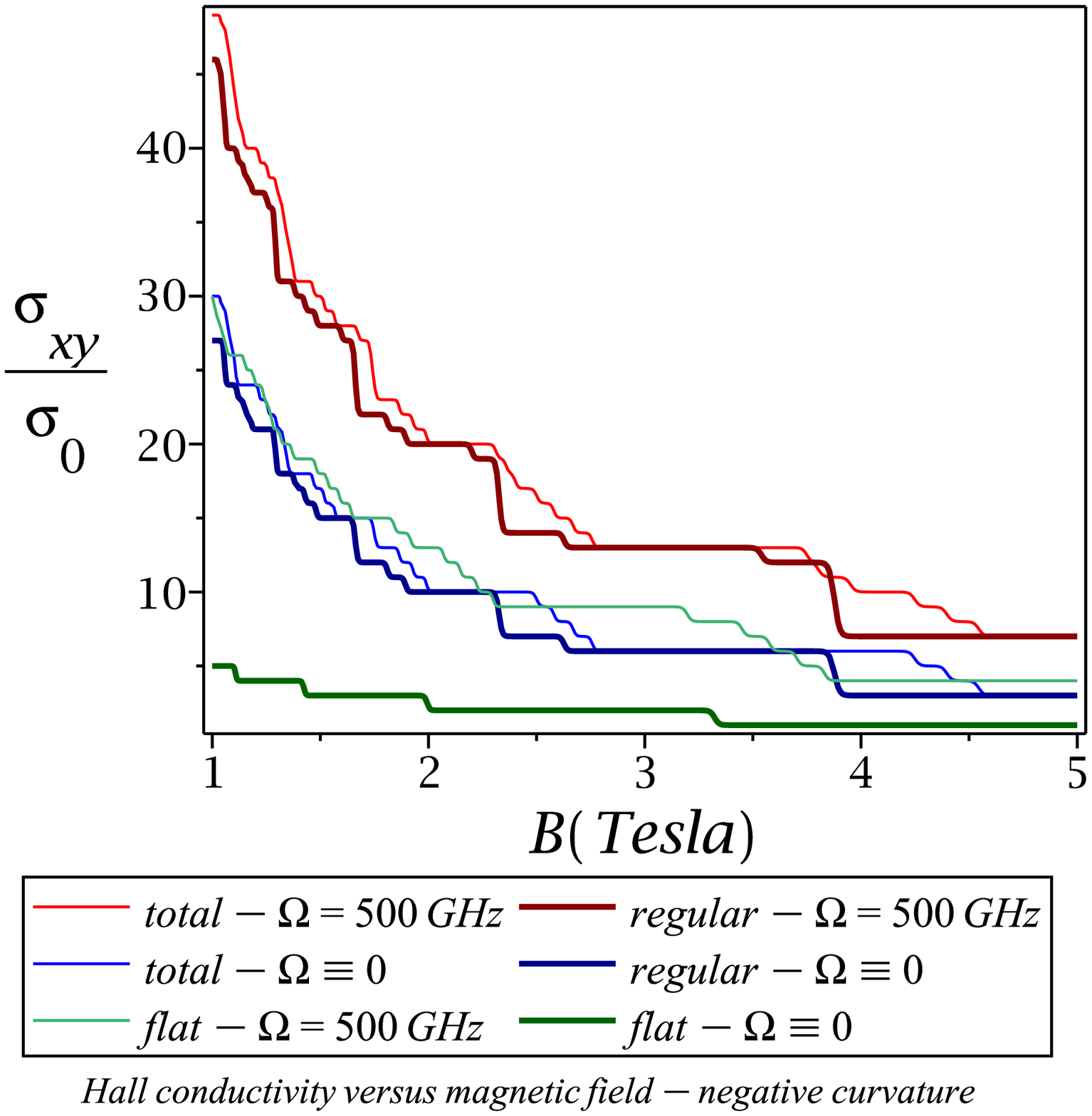}}
\caption{\small Hall steps for a rotating disk (500GHz) with disclination: {\bf (a) } $\alpha = 5/6$ and {\bf (b)} $\alpha = 7/6$. For comparison we include  the Hall steps for the stationary ($\Omega=0$) and flat  (no disclination), cases. We consider $T=0.3{\rm K}$, $\mu=100 {\rm K}$ and $\sigma_0\equiv-e^{2}/h$.} 
          \label{degraus}
\end{figure*}

\begin{figure*}[!htp]
\centering
    {\bf a}\subfloat{\includegraphics[width=0.35\linewidth]{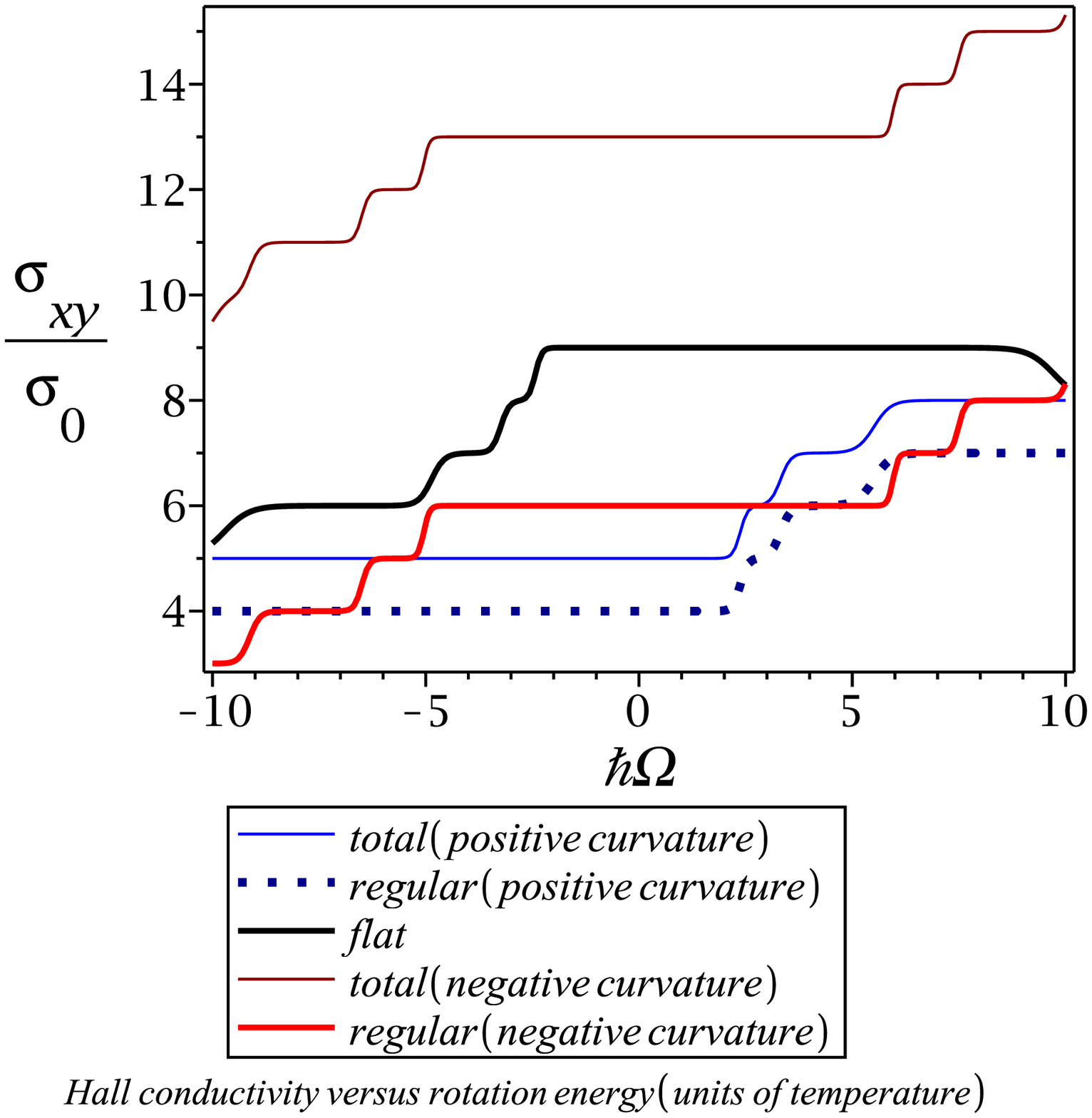}}
    {\bf b}\subfloat{\includegraphics[width=0.35\linewidth]{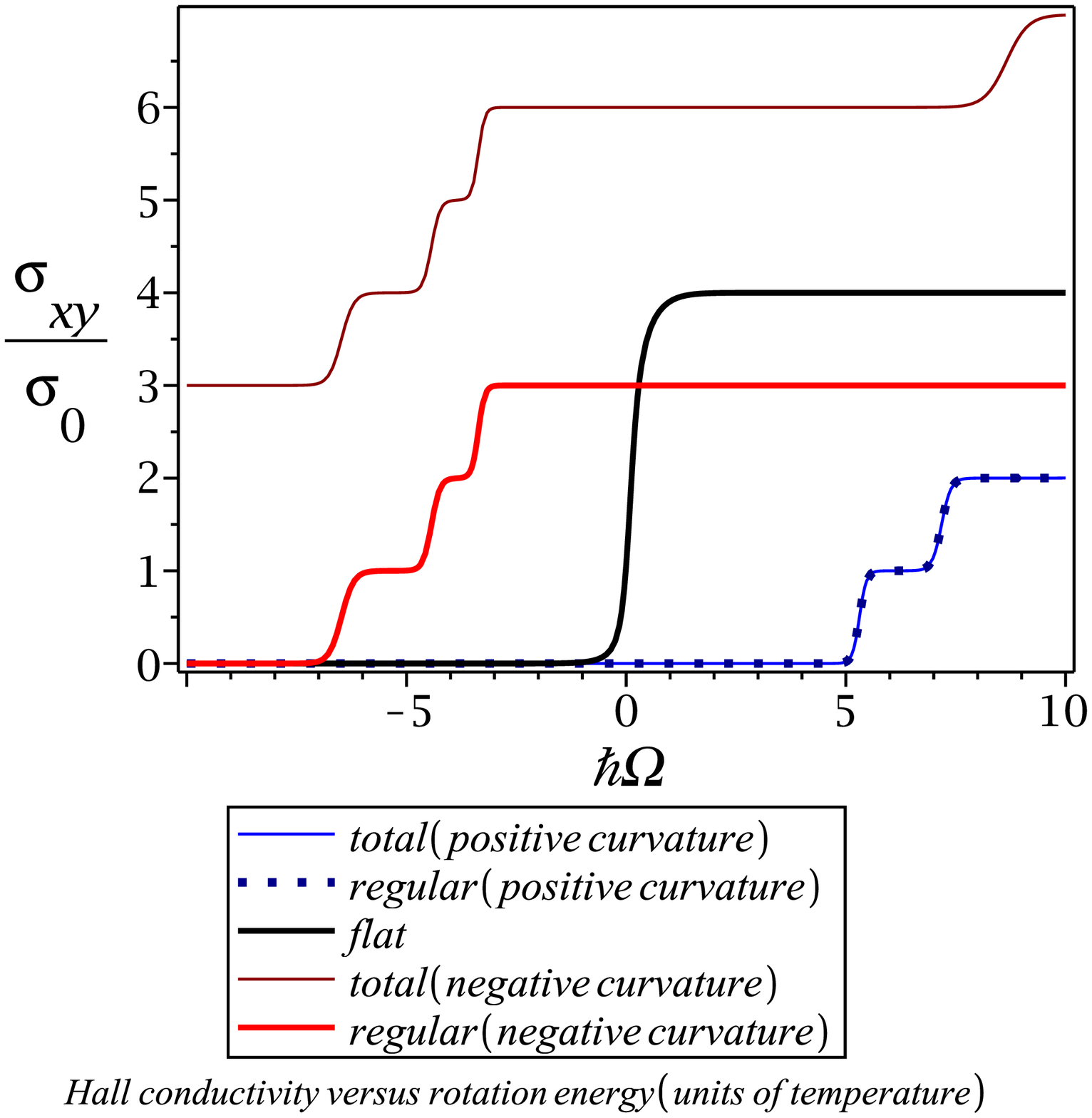}}
\caption{\small Hall conductivity versus  rotation energy for a rotating disk  with disclination: {\bf (a) } $B = 3T$ and {\bf (b)} $B=10T$. We considered $T=0.3{\rm K}$, $\mu=100 {\rm K}$ and $\sigma_0\equiv-e^{2}/h$.} 
          \label{rotacao}
\end{figure*}

The Hall conductivity versus the chemical potential  is depicted in Fig. \ref{chemical}, versus the magnetic field in Fig. \ref{degraus} and versus the rotation speed  in Fig. \ref{rotacao}. We recall that \textit{total} corresponds to the case where the effect of the defect singularity was taken into account and \textit{regular} if this is not considered.  As the magnetic field is increased while keeping the chemical potential constant, the energy levels move upwards, which is equivalent to make the highest occupied state move down. This leads to a reduction of the Hall conductivity as shown in Fig. \ref{degraus}. Considering a rotation speed of 500 GHz, for either positive ($\alpha=5/6$) or negative curvature ($\alpha=7/6$) disclination, the Hall conductivity is enhanced in the interval for the chemical potential shown in Fig. \ref{chemical}. The same occurs in the plots of the Hall conductivity versus the magnetic field shown in Fig. \ref{degraus}.   Notice also that the singularity has significant impact in the formation of the Hall steps as evidenced by the \textit{total} versus \textit{regular} curves. Figs. \ref{rotacao}.a and \ref{rotacao}.b show the  formation of plateaus due to the rotation, for different values of the magnetic field.   

\section{Conclusion}
In this paper, we investigated the influence of rotation along with a disclination on the quantized Hall conductivity of a 2D electron gas. For this purpose, we considered a rotating non-interacting planar two-dimensional electron gas in the presence of a disclination and a perpendicular uniform magnetic field. We found that, besides the joint influence of magnetic field and rotation, magnetic field and disclination, and rotation and disclination, observed in previous works, there is a simultaneous coupling between rotation, magnetic field and disclination. This interaction gives rise to a range of values for the magnetic field that does not bind the electrons into Landau levels.  The size of this region depends on the effective mass of the charge carrier, the disclination and/or the rotation rate . The most important feature of this effect is that it only exists when the three elements (magnetic field, disclination and rotation) are present simultaneously. This effect will be noted for low magnetic fields, where the Drude model can be employed to analyze the Hall effect.  This means that such effect can find applications in the context of Hall sensors, for instance.  In a  previous work, \cite{Brando}, we verified that the rotation breaks the degeneracy of the Landau levels. Here, we verified that the disclination can leverage the splitting of degenerated states. We also found the Hall conductivity for this system, which is similar to the case without rotation and disclination. Nevertheless, due to the splitting of energy levels caused by rotation and enhanced by the disclination, the number of states under the Fermi level increases, raising the Hall conductivity with respect to the previous study. As a perspective, we propose to analyze the influence of rotation and disclination together in the quantum Hall effect of novel materials such as graphene and topological insulators. Also of interest is the investigation of the influence of rotation and disclination in related phenomena like de Haas-van Alphen effect. 

\ack
We are grateful to FAPEMIG, CNPq, CAPES and FACEPE (Brazilian agencies) for financial support. 

\section*{References}
\bibliographystyle{iopart-num}
\bibliography{ref}

\end{document}